\documentclass[%
amsmath,amssymb,aps,showkeys, nofootinbib, twocolumn,superscriptaddress,longbibliography
]{revtex4-2}
\usepackage{graphicx}
\usepackage{float}
\graphicspath{{./figures/}}
\usepackage{dcolumn}
\usepackage{bm}
\usepackage{braket}
\usepackage{color}
\usepackage[normalem]{ulem}
\usepackage[colorlinks=true,linkcolor=blue,urlcolor=blue,citecolor=blue]{hyperref}

\begin{document}

\title{Defect-induced optical magnons and local magnetic correlations in MnSb$_{2}$Te$_{4}$}

\author{Dhurba~R.~Jaishi}
\affiliation{Ames National Laboratory, Ames, IA, 50011, USA}
\affiliation{Department of Physics and Astronomy, Iowa State University, Ames, IA, 50011, USA}

\author{Bing~Li}
\affiliation{Ames National Laboratory, Ames, IA, 50011, USA}
\affiliation{Department of Physics and Astronomy, Iowa State University, Ames, IA, 50011, USA}

\author{Tianxiong~Han}
\affiliation{Ames National Laboratory, Ames, IA, 50011, USA}
\affiliation{Department of Physics and Astronomy, Iowa State University, Ames, IA, 50011, USA}

\author{S.~X.~M.~Riberolles}
\affiliation{Ames National Laboratory, Ames, IA, 50011, USA}
\affiliation{Department of Physics and Astronomy, Iowa State University, Ames, IA, 50011, USA}

\author{D.~M.~Pajerowski}
\affiliation{Neutron Scattering Division, Oak Ridge National Laboratory, Oak Ridge, TN, 37831, USA}

\author{Jiaqiang~Yan}
\affiliation{Materials Science and Technology Division, Oak Ridge National Laboratory, Oak Ridge, TN, 37831, USA}

\author{R.~J.~McQueeney}
\affiliation{Ames National Laboratory, Ames, IA, 50011, USA}
\affiliation{Department of Physics and Astronomy, Iowa State University, Ames, IA, 50011, USA}

\date{\today}

\begin{abstract}
    Magnetic defect engineering offers a route to manipulate and control magnetic and electronic states in quantum materials. In the Mn(Bi,Sb)$_2$Te$_4$ family of topological magnetic insulators, Sb substitution facilitates site mixing between Mn and Sb atoms that affects magnetic order and band topology. Here we directly probe these defect-induced magnetic interactions using inelastic neutron scattering on single crystals of MnSb$_2$Te$_4$. We find that antisite mixing generates inequivalent and disordered magnetic sublattices where strong defect-induced antiferromagnetic coupling produces an optical magnon with a large gap. Semi-classical spin-dynamics simulations accurately capture magnetic excitations in the ordered and paramagnetic states and identify that linear Mn-Te-Mn bonds mediate coupling to antisite magnetic defects. 
\end{abstract}

\maketitle

\section{Introduction}
\label{sec:Intro}
    Topological insulators (TIs) possess a massless Dirac dispersion with spin-momentum locking at the surface protected by time-reversal symmetry ($\mathcal{T}$)~\cite{kane2005,kanec2005,moore2007,zhang2009top}. The breaking of $\mathcal{T}$ by the introduction of magnetism in TIs gives rise to exotic topological quantum states, such as the quantum anomalous Hall insulator and axion insulator~\cite{yu2010quantized,li2010dynamical,chang2016quantum,2017tailoring,Mogi22,bernevig2022}.  Since its discovery, the antiferromagnetic TI MnBi$_2$Te$_4$ (MBT)~\cite{Otrokov2019,Zhang19,JLi2019,Gong19} has become one of the central platforms for studying the interplay between intrinsic magnetism and nontrivial topology~\cite{Li2024review,Vyazovskaya25}, leading to the discovery of Chern and axion insulators in even and odd layered devices~\cite{Deng2020,Liu20,Lin22, qiu2025observation}  and the observation of new phenomena such as the layer Hall effect~\cite{Gao21}.  

    One early attraction to the intrinsic MBT family is that it seemingly avoided deleterious effects of magnetic disorder in dilute magnetic TIs~\cite{tokura2019magnetic} where $\mathcal{T}$ is broken by long-range coupling between random magnetic defects~\cite{YSHor2010,ZhangPRB2013,islam2023role}.  However, it is now understood that magnetic defects, caused by antisite mixing between Mn and Bi, are present and can be strongly enhanced by Sb substitution in Mn(Bi,Sb)$_2$Te$_4$~\cite{Lai2021,Liu2021,islam2023role}. These magnetic defects could be considered as imperfections that degrade the topological response due to inhomogeneity and reduction of the Dirac gap~\cite{garnica2022}.  However, defects are also recognized as an intrinsic degree-of-freedom that can be used to control magnetic order~\cite{TaitoPRB2019, Hu24, Sahoo24} and metamagnetic transitions~\cite{YSHor2010,Lai2021,Tyler_PRM}, magnetic exchange interactions~\cite{Riberolles21}, carrier density~\cite{Yan19}, and band topology~\cite{Hu21}. The sensitivity to growth conditions and subsequent thermal treatments can control details and consequences of the antisite defect configuration~\cite{Hu24}.  Defect engineering is now one of the major themes of the field, motivating extensive experimental and theoretical investigations to characterize the role of defects and disorder~\cite{Pham2019,Du2021,chen2025defect,Enk025atomic}.

    Despite extensive work on crystal growth, transport, and electronic structure, the microscopic magnetic interactions associated with antisite defects remain to be understood.  Mn defects in the Bi/Sb layer (Mn$_{\rm Sb}$) are known to have strong coupling to the main Mn layer (Mn$_{\rm Mn}$), forming defect-induced ferrimagnetism~\cite{TaitoPRB2019} with a saturation field above 50 Tesla~\cite{Lai2021}. Magnetic defects may also be responsible for the large increase in the magnetic bandwidth in Mn(Bi,Sb)$_2$Te$_4$ as measured by powder inelastic neutron scattering (INS)~\cite{Riberolles21}.  
    
    Here, we directly probe the defect-induced magnetic interactions in MnSb$_2$Te$_4$ (MST) single-crystals by INS, which reveals key details and energy scales. MST is known to have significant antisite defects ($\sim15\%$). INS data below the Neel temperature $T_{\rm N}=19$~K  find strongly renormalized spin excitations as compared to MBT, where defect concentrations are much smaller ($<3\%$). These defect-induced magnetic interactions result in a closing of the spin gap and much stiffer spin waves compared to MBT.  The observation of an optical magnon mode with a large gap is a signature of defect-induced ferrimagnetism. Classical spin-dynamics simulations using Landau-Lifshitz Dynamics (LLD) with quenched disorder accurately capture the magnetic excitations and confirm that linear Mn$_{\rm Sb}-{\rm Te}-{\rm Mn}_{\rm Mn}$ bonding generates a dominant antiferromagnetic (AFM) coupling that is much larger than the intrinsic Mn$_{\rm Mn}-{\rm Mn}_{\rm Mn}$ ferromagnetic (FM) coupling. Defect-induced, dimer-like antiferromagnetic spin correlations survive well above $T_{\rm N}$.  We conclude that antisite defects radically transform the spin dynamics of MST from those of the nominally defect-free MBT.
    \begin{figure*}[ht!]
        \centering
        \includegraphics[width=1.0\linewidth]{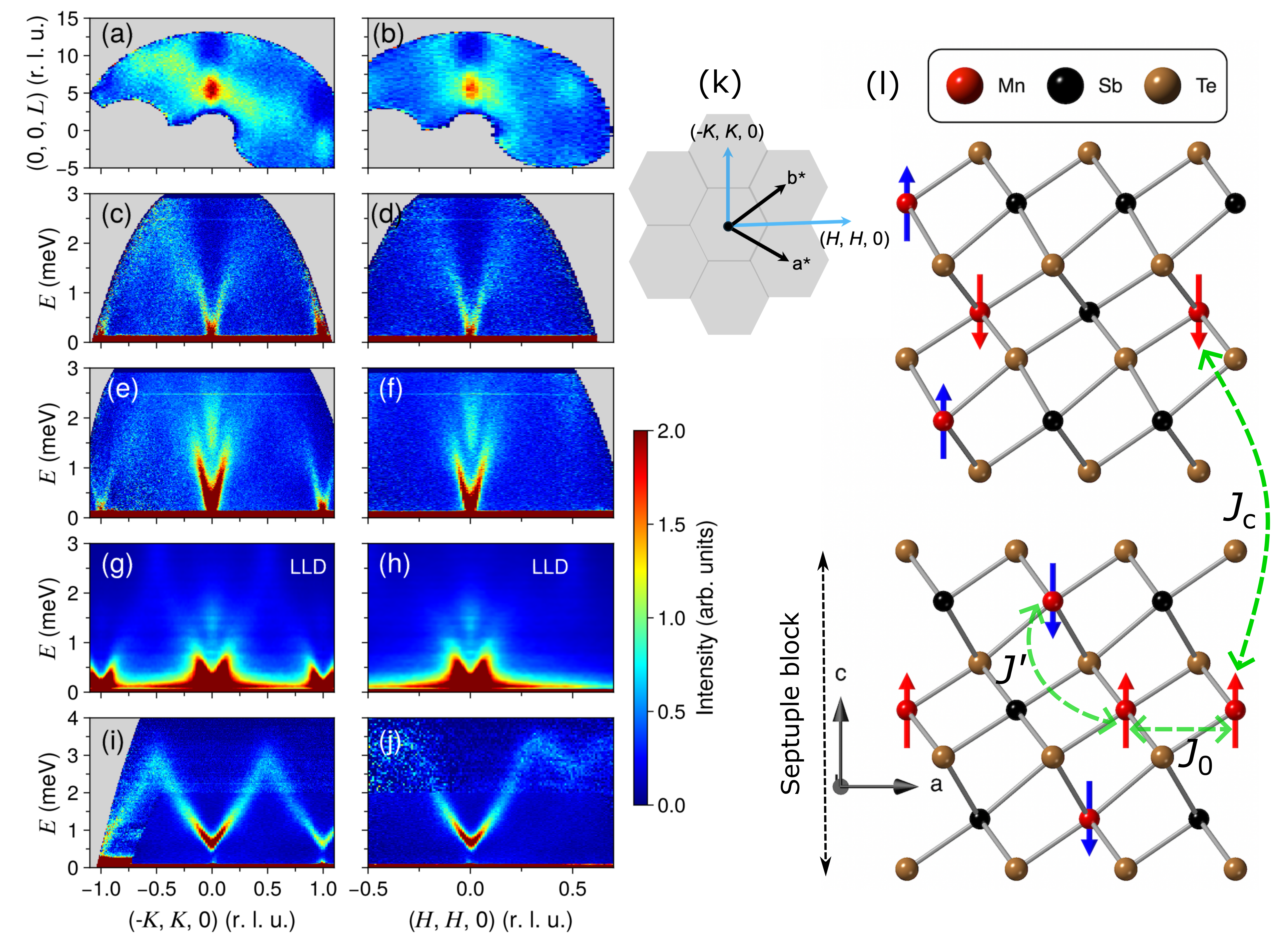}
        \caption{(a-f) INS data of AFM MnSb$_2$Te$_4$ measured at 2 K and $E_i=3.32$ meV. (a-b) Constant energy slices around $E=$ 2 meV with an energy bin width of $\Delta E= \pm 0.5$. (c-f) The in-plane spin wave dispersions. These data are integrated in the transverse direction with $H/K=$[-0.02, 0.02], $L$ = [9, 12] in (c) and (d), and $L$ = [3, 9] in (e) and (f). (g-h) The LLD simulations in the respective directions at 2 K. (i-j) The INS data of MBT measured at 2 K, where high-resolution INS data with $E_i=$~3.32 meV superposed on top of the data measured with $E_i=$~6.6 meV. The data are integrated in the $(0, 0, L)$ direction in the range of $L$ = [0, 20] for $E_i=$ of 6.6 meV and $L$ = [3, 12] for $E_i=$ of 3.32 meV. MBT INS data were reported in Ref.~\cite{BLi2021}. (k) The two-dimensional (2D) hexagonal Brillouin zone showing reciprocal lattice vectors directions and $(H, K, 0)$ scattering plane. (l) Illustration of the magnetic ordering in MnSb$_2$Te$_4$ showing magnetic impurities (red sphere) in the Sb layers and non-magnetic Sb defects (black sphere) in the Mn layers. The red and blue arrows represent moment directions, and the dashed green lines with double arrows represent the intralayer ($J_\text{0}$) and interlayer ($J'$, $J_\text{c}$) exchange interactions.  In (c-f), the sharp dispersionless feature around $\approx$ 2.5 meV is a spurious background.}
        \label{fig:disp_fig_1}
    \end{figure*}

\section{Experimental Details}
\subsection{Single crystal growth and characterization}
    The single-crystal samples of nominal MnSb$_2$Te$_4$ used in INS experiments were grown out of a Sb-Te flux, similar to the Bi-Te flux used for MnBi$_2$Te$_4$~\cite{Yan2019}. These crystals can have either FM or A-type AFM ground states, dependent on details of crystal growth and thermal treatments~\cite{Liu2021}. Here,  neutron diffraction finds a magnetic propagation vector of $\boldsymbol{\tau}=(0,0,3/2)$ representative of A-type order with $T_{\text{N}}=19$~K.  Structural refinements are consistent with the space group R$\bar{3}$m (No. 166) (same as MnBi$_2$Te$_4$) with lattice parameters  $a=b=4.2543(1)$~\AA, $c=40.899(2)$~\AA.  Refinement of neutron diffraction data finds significant Mn-Sb site mixing for the AFM sample with about $13\%$ Mn$_{\text{Sb}}$ with chemical composition (Mn$_{0.588(2)}$Sb$_{0.412(2)})$(Sb$_{0.871(2)}$Mn$_{0.129(2)}$)$_{2}$Te$_{4}$~\cite{Liu2021}.
    
\subsection{Neutron scattering}
    The INS measurements were performed at the Cold Neutron Chopper Spectrometer (CNCS) at the Spallation Neutron Source at the Oak Ridge National Laboratory. Measurements were performed on co-aligned single-crystal samples with a total mass of 503.7 mg mounted to a helium cryostat.  Two sets of measurements were performed with crystals oriented in $(H, H, L)$ and $(-K, K, L)$ horizontal scattering planes with reciprocal-space notation described below. For each orientation, measurements were performed using incident energies of $E_\text{i}$= 3.32 meV at temperatures 2 K and 30 K while the sample was rotated about the vertical axis for $\bm{Q}$, $E$ space coverage, where $\bm{Q}$ ($E$) is the momentum (energy) transfer, respectively. The INS data were collected as a function of $\bm{Q}$ and $E$ in the hexagonal reciprocal lattice units ( r. l. u.), $\bm{Q} = H{\bf a^*} + K{\bf b^*} + L{\bf c^*}$, where {\bf a}$^*$, {\bf b}$^*$, and {\bf c}$^*$ are the primitive reciprocal lattice vectors, as shown by the 2D hexagonal Brillouin zone in Fig.~\ref{fig:disp_fig_1}(k). We present the INS data in three orthogonal reciprocal directions ($H$, $H$, 0), (-$K$, $K$,0), and (0, 0, $L$). The intensities of the INS data are proportional to the spin-spin correlation function $S(\bm{Q}, E) = \sum_{\alpha, \beta}(\delta_{\alpha \beta} - \hat{Q}_\alpha \hat{Q}_\beta)S^{\alpha \beta} (\bm{Q}, E)$, where $\hat{Q}_{\alpha,\beta=x, y, z}$ are the cartesian components of the unit vector of momentum transfer $\bm{Q}$.
    
\section{Experimental Results}
\subsection{Spin waves below $T_{\rm N}$}
    INS data of AFM MST measured at $T=2$~K are shown in Fig.~\ref{fig:disp_fig_1}(a-f).  These can be compared to similar measurements performed on MBT by our group~\cite{BLi2021} and reproduced in Fig.~\ref{fig:disp_fig_1}(i, j). Fig.~\ref{fig:disp_fig_1}(a) and (b) show constant energy slices in the two crystal orientations over an energy range from 1.5$-$2.5~meV, which display strong modulation of the magnetic scattering intensity along $L$, establishing the presence of significant interlayer spin correlations within the chemical unit cell. Strong scattering is observed centered at $L_{+}=5.5$ and weak scattering at $L_{-}=11$. This corresponds to AFM spin correlations between Mn-Sb layers where $d=c/L_{-} \approx 3.7$~\AA~and already indicates a significant role of antisite mixing in the magnetic response.
    
    When we plot the dispersions in MST averaged around $L_{-}$ as shown in Fig.~\ref{fig:disp_fig_1}(e,f), a single gapless branch is observed which crosses over to heavily broadened lineshapes above $H_c\approx0.2$. This broadening is likely caused by random configurations of Mn$_{\rm Sb}$ spins and Mn$_{\rm Mn}$ vacancies. 
    At a wavelength roughly comparable to the mean distance between spins, well-defined long-wavelength collective modes of the average distribution cross over to broad, localized short-wavelength modes that are dominated by disorder. From $H_c$, we can estimate an effective spin concentration of $p_{\rm{eff}}\approx(2H_c)^2=0.16$, which is comparable to the defect concentrations measured by diffraction.
    
    Around $L_{+}$, the dispersions shown in Fig.~\ref{fig:disp_fig_1}(c,d) reveal an additional gapped branch with an onset energy of $\Delta' \approx 1.5$~meV. Given the characteristic $L$ dependence, this extra branch is consistent with large AFM coupling $\mathcal{J}'$ between Mn$_{\rm Sb}$ antisite impurities and Mn$_{\rm Mn}$ in the main layer.  In this respect, each MST trilayer behaves as a ferrimagnetic block hosting acoustic and optic spin wave branches. The optical gap $\Delta'\sim 3M_{\rm{sat}}\mathcal{J}'/g\mu_{\rm{B}}$ represents the net ferrimagnetic exchange field experienced by the minority Mn$_{\rm Sb}$ spin, the factor of 3 comes from the three neighboring bonds.  Magnetization data find $M_{\rm{sat}}\approx 1.5~\mu_{\rm{B}}$ \cite{Lai2021} and we estimate that $\mathcal{J}'\approx 0.7$~meV, which is significantly larger than the intrinsic (defect-free) magnetic couplings in MBT~\cite{BLi2020, BLi2021}.
            
    We can compare the spin excitations in MST to MBT, where the latter has a much smaller defect concentration. Previous INS data for MBT  in Fig.~\ref{fig:disp_fig_1}(i,j) show a single AFM branch that is well-defined across the Brillouin zone. In comparison to MST, the spin waves in MBT are much less stiff and possess a large anisotropy gap. Data for MBT do not display any of the features found in MST that are associated with a heavy concentration of magnetic defects.

\begin{figure}[H]
    \centering
        \includegraphics[width=1.0\columnwidth]{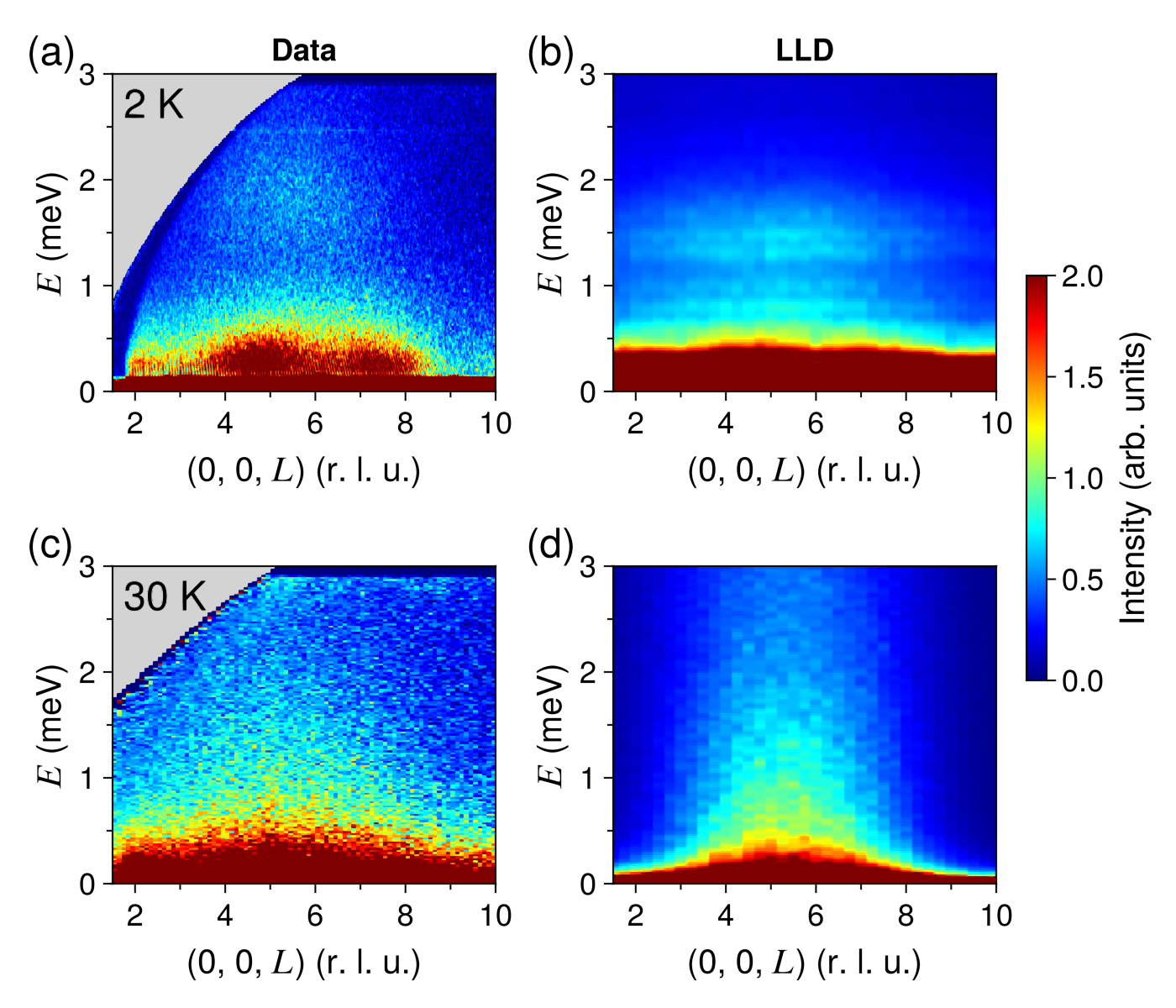}
        \caption{Spin wave excitations in the out-of-plane direction in MST. (a), (c) INS data measured at 2 K and 30 K along the $(0, 0, L)$ direction, respectively. (b), (d) LLD calculations corresponding to conditions of panels (a) and (c). The data and simulations are integrated in the transverse direction with $H/K=$[-0.02, 0.02] (r. l. u.).}
        \label{fig:mst_ool}
    \end{figure}

    The interlayer dispersion along (0, 0, $L$) for MST is shown in Fig.~\ref{fig:mst_ool}(a).  For A-type order, we expect intensity maxima to occur at $L= 3n \pm \frac{3}{2} = \frac{3}{2}, \frac{9}{2}, \frac{15}{2}, ...$ for $n=integer$ due to the staggered magnetization.  While this characteristic A-type $L$ dispersion is clearly seen in MBT~\cite{BLi2021},  it is much less obvious in MST due to its gapless character and strong contributions of Mn$_{\rm Sb}$ magnetic defects.

\subsection{Local spin correlations above $T_{\rm N}$}
\label{sec:anti-site}
\begin{figure}[H]
    \centering
        \includegraphics[width=1.0\columnwidth]{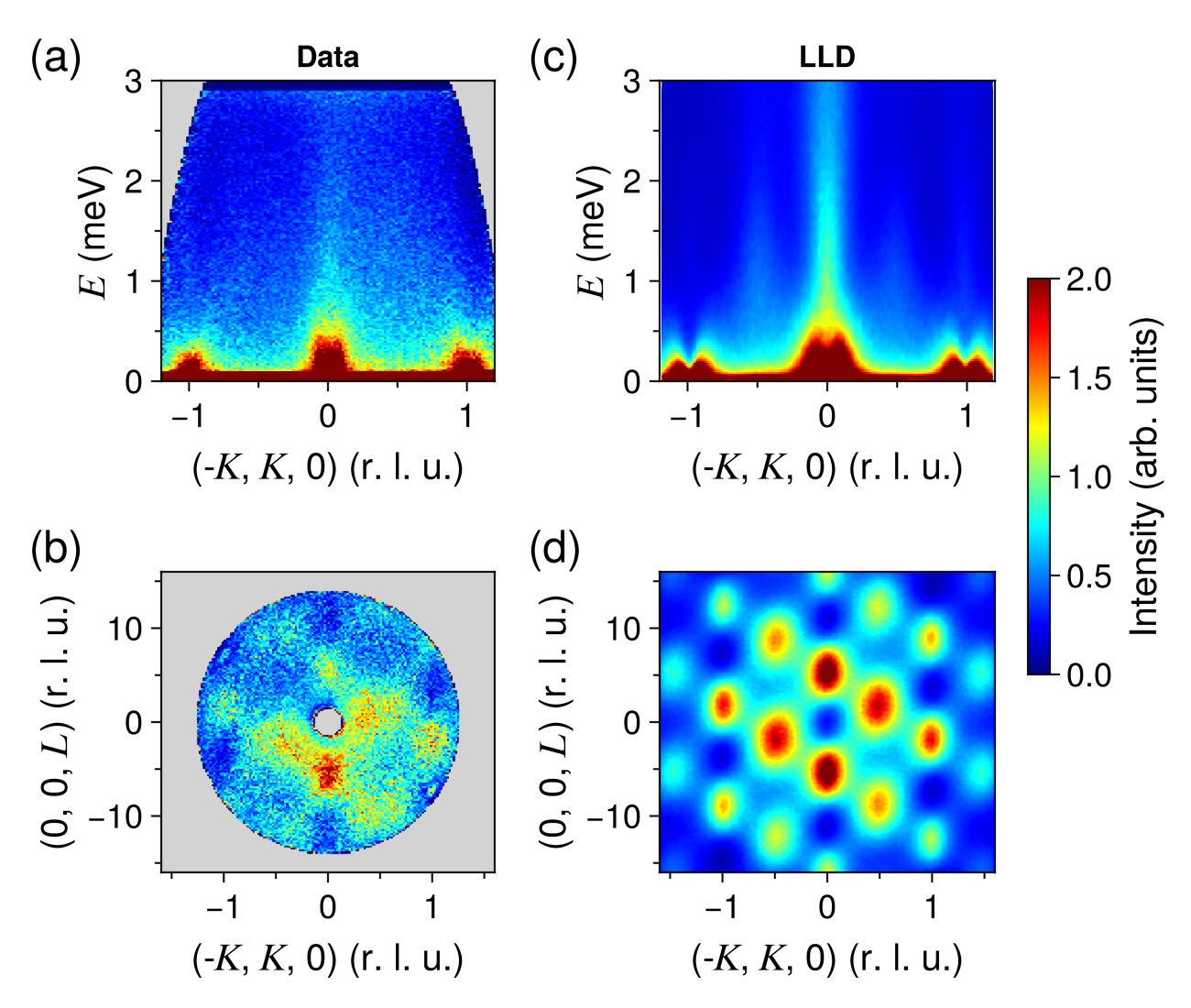}
        \caption{Magnetic dimer formation above $T_{\rm N}$. (a) INS data measured along the in-plane direction at $T = 30$~K. The data are integrated in the transverse directions with $H=$[-0.1, 0.1] and $L=$[3, 9] (r. l. u.). (b) Constant energy cut around $E=$ 1 meV with an energy bin width of $\Delta E= \pm 0.1$. (c-d) LLD simulation of INS scattering intensity at $T=30$~K with sampling corresponding to panels (a) and (b), respectively.}
        \label{fig:disp_fig_2}
    \end{figure}

    The large optical magnon gap $\Delta'$ indicates a strong coupling between Mn$_{\text{Sb}}$ and Mn$_{\text{Mn}}$ with an energy scale that is larger than $k_{\rm B}T_{\rm N}$. Indeed, the signature of coupling to Mn$_{\text{Sb}}$ defects persists above $T_{\text{N}}=19$~K. Figure~\ref{fig:disp_fig_2}(a) shows INS data measured at $T=30$~K along the in-plane direction, indicating residual spin correlations near the Brillouin zone center.  We obtain information about the spatial correlations in the ($-K,K,L$) plane from a constant energy cut around $E=$ 1 meV with an energy bin width of $\Delta E= \pm 0.1$, as shown in Fig.~\ref{fig:disp_fig_2}(b). The maxima observed at (0, 0, 5.5) indicate that strong AFM coupling of moments in the Mn and Sb layers persists above $T_{\rm N}$.
    
    The full ${\bm Q}$ dependence of the spin correlations in Fig.~\ref{fig:disp_fig_2}(b) displays a pattern that resembles the structure factor of an isolated magnetic dimer (proportional to $\cos{(\bm Q \cdot \bm R)}$, where ${\bm R}$ is a vector defining the intradimer distance) \cite{Furrer2013}.  Comparison between the calculated dynamical structure factor of the AFM dimers formed by interlayer nearest (Mn$_{\text{Mn}}$-Mn$_{\text{Sb}}$) and next-nearest neighbor (Mn$_{\text{Mn}}$-Te-Mn$_{\text{Sb}}$) bond vectors are shown in the Appendix in Fig.~\ref{fig:dimer}. The next-nearest-neighbor structure factor shows good agreement with the measured structure factor shown in Fig.~\ref{fig:disp_fig_2}(b). This establishes that strong AFM dimers are formed by pairs of Mn atoms that are connected by almost linear Mn-Te-Mn bonds within the septuple block. A similar conclusion regarding AFM Mn-Mn dimer formation in Mn-doped Sb$_2$Te$_3$~\cite{islam2023role} suggests that this is a common feature of quintuple and septuple layers in the broad family of topological insulators.
    
    The observed structure factor also supports the argument that magnetic defects (Mn$_{\rm{Sb}}$) are positioned in the Sb site rather than randomly lying in interstitial defects. However, no distinct levels associated with the dimer spin-state transitions are observed (see eg. Ref.~\cite{Vaknin20}), but only a continuum in energy as shown in Fig.~\ref{fig:disp_fig_2}(a). 

\section{Analysis of INS data}
\subsection{Spin Hamiltonian}
\label{sec:mag_ham}
    We analyze the neutron scattering data using calculations from a minimal spin Hamiltonian.  Based on the qualitative analysis above, we can define a spin Hamiltonian for MST that involves magnetic defect disorder, including Mn layer vacancies with concentration $1-p$ and antisite Mn$_{\rm Sb}$ spins with concentration $q$.  We simplify the interactions between spins in Mn layers to include FM nearest-neighbor intralayer coupling ($\mathcal{J}_0<0$) and AFM interlayer coupling across septuple layer blocks ($\mathcal{J}_c>0$). We introduce a strong AFM coupling ($\mathcal{J}'>0$) between Mn$_{\rm Sb}$ and Mn$_{\rm Mn}$ spins within the same septuple block and along the Mn-Te-Sb bond.  These interactions are shown graphically in Fig.~\ref{fig:disp_fig_1}(l). The spin Hamiltonian can be written as, 
    \begin{align}
       \mathcal{H}&=\mathcal{J}_0 \sum_{\langle i, j\rangle} p_ip_j{\mathbf{S}_{i}} \cdot {\mathbf{S}_{j}} + \mathcal{J}_c\sum_{\langle i, j\rangle} p_ip_j{\mathbf{S}_{i}} \cdot {\mathbf{S}_{j}} \nonumber \\ 
       &+ \mathcal{J}' \sum_{\langle\langle i, k\rangle\rangle}p_iq_k{\mathbf{S}_{i}} \cdot {\mathbf{S}_{k}} + D\big[\sum_{i} p_i({S^{z}_{i})^{2}} + \sum_{k} q_k({S^{z}_{k})^{2}}\big]
    \end{align}
    where $i,j$ labels the Mn$_{\rm Mn}$ spins and $k$ labels Mn$_{\rm Sb}$ spins. The angular brackets $\langle ... \rangle$ and $\langle\langle ... \rangle\rangle$ denote nearest and next-nearest neighbor sites, respectively. We assume that the single-ion anisotropy $D$ is the same for Mn spins at Mn$_{\rm Mn}$ and Mn$_{\rm Sb}$ sites.  Given the absence of an observable spin gap, $D$ is assumed to be very small for MST and we set it to zero. Here, $p_i =$ 0 or 1 represents the defect disorder distribution in the Mn layer, and $\sum_i^N p_i/N=p$. A similar definition is used for the Sb layer, $q_k$. 

\subsection{Effective linear spin-wave theory model}
\label{sec:LSWT}
    We start by using an effective linear spin-wave theory (LSWT) model to extract the key parameters of the spin Hamiltonian. Unlike the LLD approach below, LSWT is not readily applicable to magnetically disordered systems.  However, an effective model can be developed which treats the Mn and Sb layers as homogeneous, but with reduced average moments $S_{\rm Mn} = pS$ and $S_{\rm Sb} = qS$.  We assume that $D=0$ and $\mathcal{J}_c=0$, resulting in a tri-layer model of the septuple block that can be used to extract the effective strength of magnetic interactions. Using LSWT, the energy dispersion of excitations in the tri-layer model is analytically derived (see Appendix~(\ref{sec:model_ana}) for details). This approach allows for fitting of the data (which is more difficult with LLD) to extract effective values for the spin Hamiltonian parameters.
    \begin{figure}[H]
    \centering
        \includegraphics[width=1.0\columnwidth]{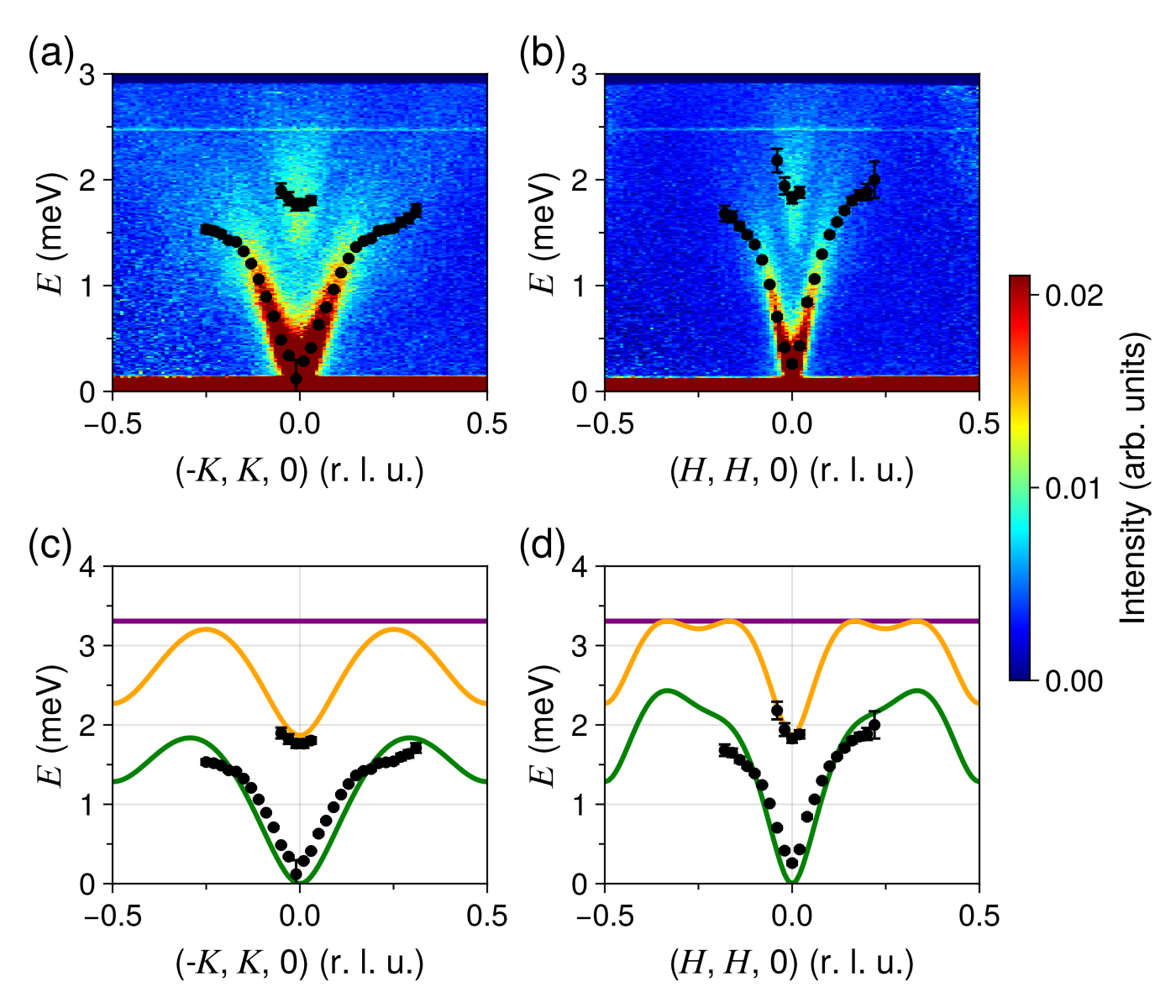}
        \caption{(a-b) INS data measured along $(-K, K, 0)$ and $(H, H, 0)$ directions at 2 K, respectively, overplotted with the extracted peak centers from fitting. The data are integrated in the $(0, 0, L)$ direction in the range of $L$ = [3, 9] (r. l. u.). (c-d) Calculated spin wave dispersion using the minimal tri-layer model.}
        \label{fig:model}
    \end{figure}
    To fit the data, line cuts are made at constant $\bm{Q}$ positions and the resulting lines are fitted to a Gaussian lineshape. The peak centers are then extracted and fitted with the analytical model dispersion shown in Figs.~\ref{fig:model}(a-d).  Taking the chemical composition of the AFM samples determined from the neutron diffraction experiment, (Mn$_{0.588(2)}$Sb$_{0.412(2)}$)(Sb$_{0.871(2)}$Mn$_{0.129(2)}$)$_{2}$Te$_{4}$ with $p=0.588$ and $q=0.129$ we approximate that $S_{\rm Mn}\approx 1.47$ and $S_{\rm Sb}\approx 0.32$.
    
    From fitting to the peak centers, we get the strength of the FM interaction $\mathcal{J}_0=-0.075(1)$~meV and AFM interaction $\mathcal{J}'=0.75(3)$~meV. The value for $\mathcal{J}_0$ is similar to that found in MBT~\cite{BLi2020,BLi2021} and MnBi$_4$Te$_7$~\cite{BLi2025}. The value for $\mathcal{J}'$ is also in agreement with the value estimated from magnetization in high field~\cite{Lai2021} and extracted from INS data on dilute Mn substitution in Sb$_2$Te$_3$ aTI~\cite{islam2023role}. Thus, the defect-induced AFM coupling $\mathcal{J}'$ is about 10 times strong as the dominant FM interaction $\mathcal{J}_0$.

    One open question is the origin of the much larger spin stiffness of the acoustic mode in MST as compared to MBT. Using the analytical forms for the spin dispersions, we can investigate the role that defect interactions play in the stiffness. The acoustic spin stiffness ($A$) is given by $E_1 \approx AH^2$ at small $H$ for propagation along ${\bm Q} = H{\bf a}^*$.  Using our analytical expressions, we find that $A/A_0 = (3p^2+8pq|\mathcal{J}'/\mathcal{J}_0|)/6(p-2q)$ where $A_0$ is the spin stiffness of defect-free MBT.  Generally, $A/A_0\geq 1$ and for $p=0.588$, $q=0.13$, and $|\mathcal{J}'/\mathcal{J}_0| \approx 10$, we find that $A/A_0 \approx 3.6$.  This predicted increase in spin stiffness due to defects is in good accordance with experimental data, where we estimate that $A/A_0 \approx 3$. Thus, the large increase in spin stiffness is also driven by magnetic defects.
 
\subsection{Classical spin dynamics}
\label{sec:LLD_sim}
    To compute the dynamical spin structure factor of MST while accounting for a random defect configuration, we performed simulations using stochastic LLD implemented in the \textsc{sunny.jl}~\cite{Dahlbom2025}. The Landau Lifshitz equation, for the spin dipole, ${\bf S}_j$ with phenomenological damping parameter $\lambda$, and with noise term, ${\bm \xi}_j$ can be written as
    \begin{equation}
        \frac{d{\bf S}_j}{dt}=-{\bf S}_j\times \Big({\bm \xi}_j +\frac{\partial \mathcal{H}}{\partial {\bf S}_j}-\lambda {\bf S}_j\times\frac{\partial \mathcal{H}}{\partial {\bf S}_j}\Big) 
    \end{equation}
    The simulations were performed using a 40 $\times$ 40 $\times$ 4 supercell of the chemical unit cell (the magnetic unit cell is twice the chemical unit cell along the $c$-axis), with periodic boundary conditions. We randomize throughout the supercell with vacancy concentration of $1-p=0.42$ in the Mn layers and $q=0.13$ Mn$_{\rm Sb}$ impurity concentration to represent pure antisite mixing with $1-p \simeq 3q$, which is close to the actual chemical composition of our MST samples. We used $\mathcal{J}_0=-0.075$~meV, $\mathcal{J}'=10\times \mathcal{J}_0$, $\mathcal{J}_c=0.002$ meV, and $D=-0.002$ ~meV for LLD simulations.
            
    The spin system was first thermalized by performing 10,000 Langevin time steps with a damping constant 0.1 and a time step $\Delta t$ = 0.02 meV$^{-1}$. After thermalization, 24 spin configurations were sampled using the stochastic Landau-Lifshitz equation, with each configuration decorrelated by 12,000 Langevin time steps between the sampling. Finally, the spin trajectories are spatiotemporal Fourier-transformed to obtain $S({\bm Q},E)$. 
    
    The calculated order parameter, shown in Appendix~\ref{sec:ord_param} is comparable with the measured $T_{\rm N}$. Figs.~\ref{fig:disp_fig_1}(e-f) show simulated spin wave dispersions using LLD at $T=2$~K in the respective directions. The LLD calculations do a reasonably good job of capturing the main features that distinguish the spin dynamics of MST and MBT.  They capture the development of a ferrimagnetic optical branch where the optical gap requires a strong defect-driven exchange of $\mathcal{J}'=$ 0.75 meV.  We also find that $\mathcal{J}'$ generates a larger effective spin stiffness when compared to MBT, in agreement with the effective LSWT calculations. Finally, the random disorder of vacancies and defects does result in short-wavelength broadening that mimics the observed data.  LLD calculations of broad and modulated interlayer $(0, 0, L)$ dispersion shown in Fig.~\ref{fig:mst_ool}(b) show good agreement with the observed data shown in Fig.~\ref{fig:mst_ool}(a). 
    
    LLD calculations were also performed above $T_{\rm N}$ at a simulation temperature of 30 K and are shown in Fig.~\ref{fig:disp_fig_2}(c) and (d).  These calculations also enforce that Mn$_{\rm Sb}$ defects remain strongly coupled to the Mn layers.  LLD calculations in Fig.~\ref{fig:mst_ool}(d) show that washed-out spin correlations in the out-of-plane direction still remain peaked at $L=5.5$, in agreement with experimental data. Given the similarity to simple dimer structure factor calculations shown in Fig.~\ref{fig:disp_fig_2}(d), the LLD calculations support the existence of dimer-like defect-bound states between Mn$_{\rm Sb}$ and Mn$_{\rm Mn}$ spins above $T_{\rm N}$.

\section{Conclusions}
\label{sec:conclusions}
    We present INS measurements on single crystals of the A-type AFM MST, which host strong Mn/Sb antisite mixing. In MBT, the septuple blocks nominally contain a single FM layer and display a single, well-defined spin wave branch. In MST, antisite mixing generates inequivalent and disordered magnetic sublattices where strong defect-induced AFM coupling leads to ferrimagnetic septuple blocks. Thus, MST provides an ideal system for studying the role of magnetic defects and magnetic vacancy disorder in spin dynamics. Using INS, we observe key features that are driven by magnetic disorder, including a crossover from long-wavelength collective modes to broad, short-wavelength excitations dominated by disorder. We also observe a defect-driven optical magnon branch where the optical gap, originating from the net ferrimagnetic exchange field within the septuple block, is a signature of defect-driven magnetic interactions.

    Our analysis finds that the defect-induced coupling is strongly AFM, as evidenced by the relatively large optical gap. The ${\bm Q}$ dependence of the scattering conclusively identifies that this coupling occurs through nearly linear Mn$_{\rm Mn}$-Te-Mn$_{\rm Sb}$ bonds that are optimized for large AFM coupling via the Goodenough-Kanamori rules~\cite{Goodenough,Kanamori}.  This bonding configuration is quite general, and linear Mn-Te-Mn bonding has been demonstrated to form strong AFM dimer singlets in Mn-substituted dilute magnetic TIs based on SnTe \cite{Vaknin20} and Sb$_2$Te$_3$~\cite{islam2023role}.

    The dominant defect-induced exchange interaction is nearly an order-of-magnitude larger than the intrinsic FM exchange of the Mn layer. Consequently, antisite defects should be regarded not simply as sources of disorder but as an intrinsic magnetic subsystem that reshapes the collective spin dynamics. 

    As the defect concentration increases, interlayer interactions develop between magnetic impurities across the van der Waals gap separating the septuple blocks. Previous work indicates that this defect-induced interlayer interaction is FM and competes with the AFM Mn$_{\rm Mn}$-Mn$_{\rm Mn}$ interaction ($\mathcal{J}_c$)~\cite{TaitoPRB2019,Liu2021,islam2023role}. This competition can favor long-range ferrimagnetism that breaks the global $\mathcal{T}$. In our data, the $(0, 0, L)$ dispersion contains information about this competition, but higher resolution data is needed to resolve these competing interactions.

    Overall, these results demonstrate that antisite defects create a second magnetic subsystem whose exchange interactions dominate the intrinsic magnetic energy scales and fundamentally change the magnetic Hamiltonian of MBT-based magnetic TIs. The antisite concentration can be controlled through growth and annealing and these results identify microscopic magnetic interactions that can be tuned through defect engineering.  Since magnetic order underpins the topological phases realized in the MBT family, understanding how antisite defects modify the microscopic magnetic Hamiltonian is an essential step toward controlling their topological properties through defect engineering.
    
\section*{Data Availability}
\label{sec:data_availability}
    The data that support the findings of this article are openly available here \cite{IPTS}

\section*{Acknowledgements}
\label{sec:acknowledgements}
    This work is supported by the U.S. Department of Energy (U.S. DOE), Office of Basic Energy Sciences (BES), Division of Materials Sciences and Engineering through the Ames National Laboratory under Contract No. DE-AC02-07CH11358. This research used resources at the Spallation Neutron Source, a DOE Office of Science User Facility operated by the Oak Ridge National Laboratory. The beam time was allocated to CNCS on proposal numbers IPTS-24428 and IPTS-27080. 
    
\setcounter{figure}{0}
\renewcommand{\thefigure}{A\arabic{figure}}
\appendix

\section{LLD simulation of $T_{\rm N}$}
\label{sec:ord_param}
    \begin{figure}[H]
    \centering
        \includegraphics[width=1.0\columnwidth]{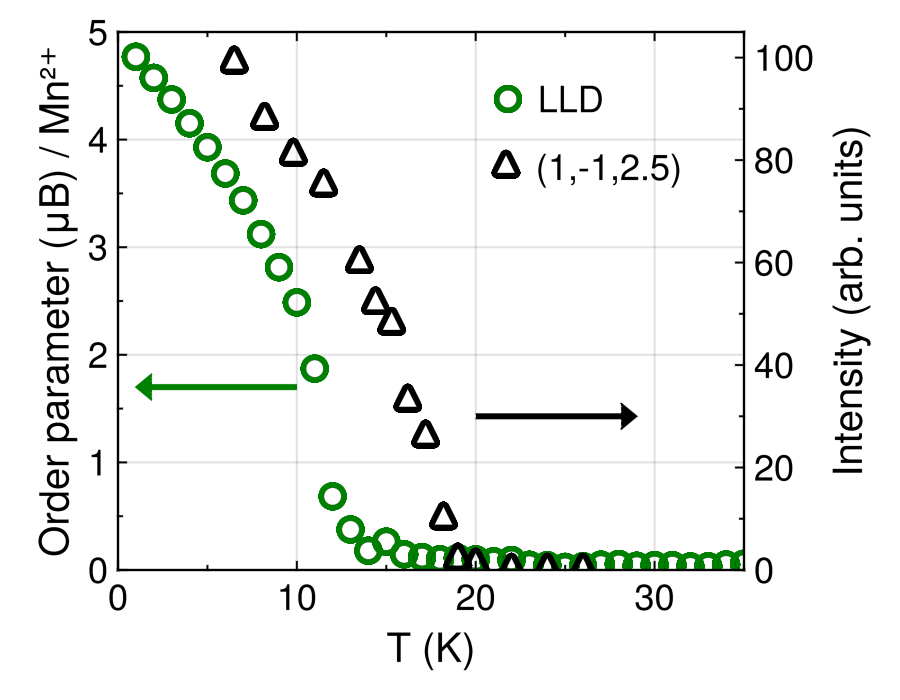}
        \caption{Calculated order parameter as a function of temperature. The data points (triangle) are neutron diffraction measurements at magnetic Bragg peak (1, -1, 2.5) reproduced from Ref.~\cite{Liu2021}.}
        \label{fig:ord_param}
    \end{figure}
    The calculated order parameter a function of temperature in dilute-MnSb$_2$Te$_4$ is shown in Fig.~\ref{fig:ord_param} using the model Hamiltonian explained in the main text. The simulations were performed using LLD.
    
\section{AFM dimer structure factor}
    \begin{figure}[H]
    \centering
        \includegraphics[width=1.0\columnwidth]{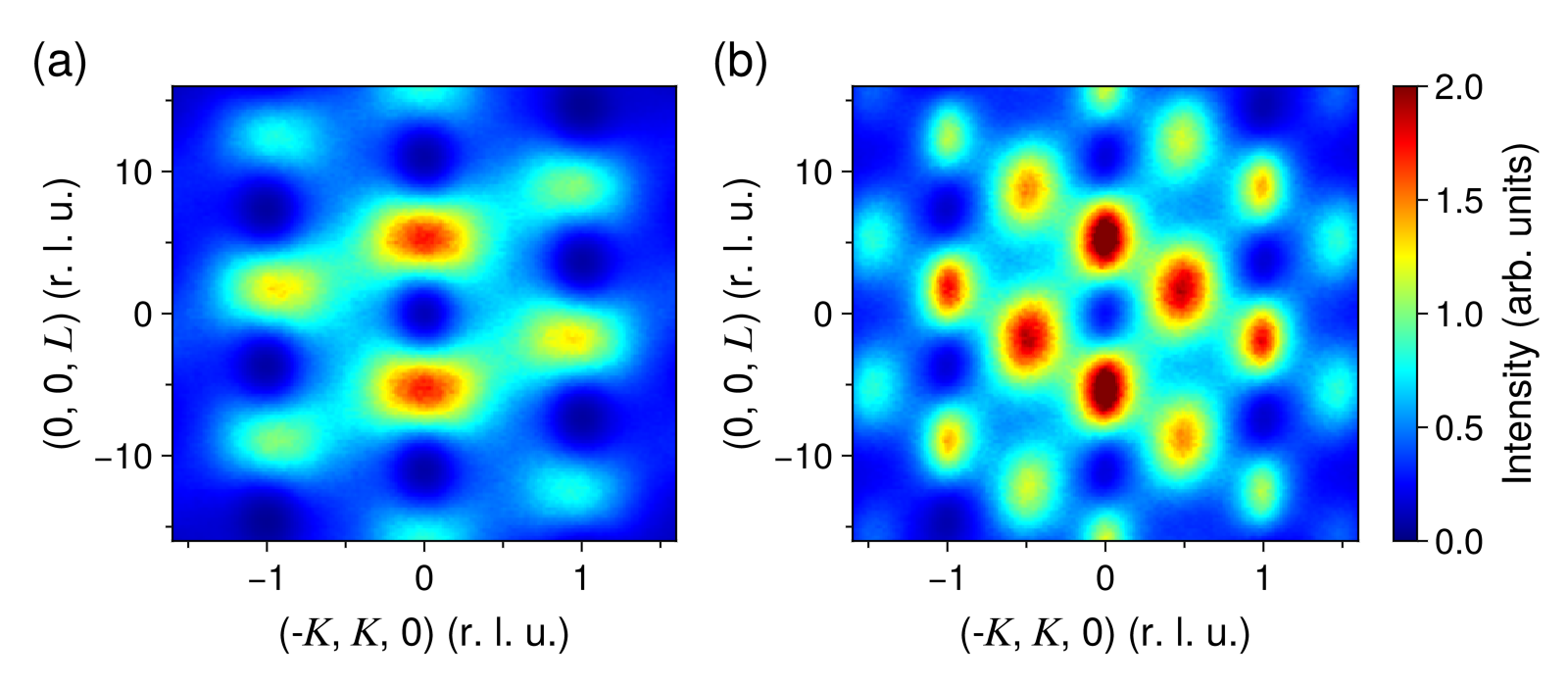}
        \caption{Magnetic dimer formation. Calculated INS scattering intensity of magnetic dimer formed via (a) nearest-neighbor, (b) next-nearest neighbor bond at $T=$~30 K centered around $E=$~1 meV with an energy bin width of $\Delta E=\pm$~0.1.}
        \label{fig:dimer}
    \end{figure}
    Figure~\ref{fig:dimer} shows the calculated magnetic dimer structure factor. Fig.~\ref{fig:dimer}(a) and (b) show the calculated dimer structure for nearest-neighbor and next nearest-neighbor bonds with $J'$, respectively. The magnetic dimer formed between Mn$_{\rm{Mn}}$ and Mn$_{\rm{Sb}}$ along the linear Mn-Te-Mn bond shows good agreement with the experimental observation. Here, ${\bm Q}$ dependence of the structure factor is an unambiguous fingerprint of the interaction geometry.
    
\section{Minimal tri-layer analytical model}
\label{sec:model_ana}
    The spin wave dispersions presented in Figs.~\ref{fig:model}(a) and (b) are heavily broadened. To extract the strength of the magnetic interactions, line cuts are made at constant $\bm{Q}$ positions and fit to a Gaussian lineshape. The peak centers are then extracted (also overplotted in Fig.~\ref{fig:model}(a) and (b)) and fitted to the analytical model dispersion, which we will develop in the following.
    
    We propose a minimal analytical model to describe the in-plane spin wave dispersion by considering only the Mn$_{\text{Sb}}$-Mn$_{\text{Mn}}$-Mn$_{\text{Sb}}$ tri-layer block within a septuple block as shown in Fig.~\ref{fig:disp_fig_1}(h). The tri-layer model consists of three hexagonal close-packed layers with lattice vectors $\bm{a}_{1}=(a,0,0)$ and $\bm{a}_{2}=(-\frac{a}{2}, \frac{\sqrt{3}a}{2} ,0)$. Mn$_{\rm{Mn}}$ positioned at origin, and there are six nearest neighbor within the layer with bond vectors; $\bm{l}_{1,2}=(\pm a, 0,0)$, $\bm{l}_{3,4}=(\pm \frac{a}{2}, \pm \frac{\sqrt{3}a}{2},0)$, and $\bm{l}_{5,6}=(\mp \frac{a}{2}, \pm \frac{\sqrt{3}a}{2},0)$. There are six Mn$_{\text{Sb}}$ interlayer second nearest neighbor; three of them are above the main Mn layer with bond vectors; $\bm{d}^u_1=(a, \frac{a}{\sqrt{3}}, (z-\frac{1}{3})c)$, $\bm{d}^u_2=(-a, \frac{a}{\sqrt{3}}, (z-\frac{1}{3})c)$, and $\bm{d}^u_3=(0, -\frac{2a}{\sqrt{3}}, (z-\frac{1}{3})c)$, and three below the main Mn layer, $\bm{d}^d_1=-\bm{d}^u_1$, $\bm{d}^d_2=-\bm{d}^u_2$ and $\bm{d}^d_3=-\bm{d}^u_3$. Each layer is FM, but the Sb layers above and below are AFM coupled with the central layer. The central layer has a reduced spin $S_A=pS$, and nearest-neighbor FM intralayer exchange coupling $\mathcal{J}_0<0$, and top and bottom layers have spin $S_B=qS$ and couple to the central layer by the AFM $\mathcal{J}'>0)$. We ignored the intralayer coupling in the top and bottom layers (suitable for the dilute occupancy of Mn in the top and bottom Sb layers).
    
    The minimal Hamiltonian for a tri-layer model can be written as, 
    \begin{equation}
        \mathcal{H}=\mathcal{J}_0 \sum_{\langle i, j\rangle} {\mathbf{S}_{A,i}} \cdot {\mathbf{S}_{A,j}} + \mathcal{J}' \sum_{\langle i, j\rangle} {\mathbf{S}_{A,i}} \cdot {\mathbf{S}_{B,j}}
        \label{Ham_sup}
    \end{equation}
    To calculate excitation energy for spin waves, we write the spin operators in terms of bosonic creation and annihilation operators ($S^{\pm}=S^x\pm iS^y$) on every magnetic sites keeping only the lowest order of boson operators using Holstein-Primakoff approximation:
    \begin{equation*}
    \begin{aligned}
        S^{+}_{A} &=\sqrt{2S_A}~a, & S^{u+}_{B}&=\sqrt{2S_B}~b^{\dagger}, & S^{l+}_{B}&=\sqrt{2S_B}~c^{\dagger} \\
        S^{-}_{A} &=\sqrt{2S_A}~a^{\dagger}, & S^{u-}_{B} &=\sqrt{2S_B}~b, & S^{l-}_{B}&=\sqrt{2S_B}~c \\
        S^{z}_{A} &=S_A-~a^{\dagger}a, & S^{uz}_{B} &=-S_B+b^{\dagger}b, & S^{lz}_{B}&=-S_B+c^{\dagger}c
    \end{aligned}
    \end{equation*}
    To diagonalize Hamiltonian~(Eq.~\ref{Ham_sup}), bosonic operators can be expressed as 
    \begin{align*}
        a^j_i&=\frac{1}{\sqrt{N}}\sum_\text{\bf q} \text{e}^{i\text{\bf q}\cdot\text{\bf r}_j} a_q~, & a^{j{\dagger}}_i&=\frac{1}{\sqrt{N}}\sum_\text{\bf q} \text{e}^{-i\text{\bf q}\cdot\text{\bf r}_j} a^\dagger_q
    \end{align*}  
    with Hamiltonian reads,
    \begin{equation}
    \begin{aligned}
        \mathcal{H} = &\mathcal{E}_0 + \sum_{\bf q}A_{\bf q}a_{\bf q}^{\dagger}a_{\bf q} + B_{\bf q}(b_{\bf q}^{\dagger}b_{\bf q} + c_{\bf q}^{\dagger}c_{\bf q})\\ &+C_{\bf q}(a_{\bf q}b_{-{\bf q}} + a_{\bf q}^{\dagger}b^{\dagger}_{-{\bf q}})+D_{\bf q}(a_{\bf q}c_{-{\bf q}} + a_{\bf q}^{\dagger}c^{\dagger}_{-{\bf q}})
    \end{aligned}
    \end{equation}
     where, $A_\mathbf{q}=2S_A\mathcal{J}_0\{\cos{(\mathbf{q} \cdot \mathbf{a}_1)}+\cos{(\mathbf{q} \cdot \mathbf{a}_2)}+\cos{(\mathbf{q} \cdot (\mathbf{a}_1+\mathbf{a}_2))}-3\}+6S_B\mathcal{J}'$, $B_\mathbf{q}=3S_A\mathcal{J}'$,    $C_\mathbf{q}=\sqrt{S_AS_B}\mathcal{J}'\sum_je^{i\mathbf{q} \cdot \mathbf{d}^u_j }$, and $D_\mathbf{q}=C_{-\mathbf{q}}$.

    Finally, using Bogoliubov linear transformation, we find three positive eigenvalues of $\mathcal{H}$;
    \begin{equation}
    \begin{aligned}
        E_{1} &= \frac{1}{2}\bigg(\alpha_{\bm{q}} + \sqrt{\alpha_{\bm{q}}^2 +\beta_{\bm{q}}}\bigg) \\
        E_{2} &=\frac{1}{2}\bigg(-\alpha_{\bm{q}} + \sqrt{\alpha_{\bm{q}}^2 +\beta_{\bm{q}}}\bigg) \\
        E_{3} &= 3S_{A}\mathcal{J}'
    \end{aligned}
    \end{equation}
    where, $\alpha_{\mathbf{q}}=A_{\mathbf{q}}-B_{\mathbf{q}}=3\mathcal{J'}(2S_B-S_A)-6S_A\mathcal{J}_0f(\mathbf{q})$, $\beta_\mathbf{q}=4A_{\mathbf{q}}B_{\mathbf{q}}-8|C_{\mathbf{q}}|^2=-72S_A^2\mathcal{J}_0\mathcal{J'}f(\mathbf{q})+48S_AS_B\mathcal{J'}^2f(2\mathbf{q})$, and $f(\mathbf{q}) = 1 - \frac{1}{3}[\cos{(\mathbf{q} \cdot \mathbf{a}_1)}+\cos{(\mathbf{q} \cdot \mathbf{a}_2)}+\cos{(\mathbf{q} \cdot (\mathbf{a}_1+\mathbf{a}_2))}]$.
    
    Taking the chemical composition of the AFM samples determined from the neutron diffraction experiment, (Mn$_{0.588(2)}$Sb$_{0.412(2)}$)(Sb$_{0.871(2)}$Mn$_{0.129(2)}$)$_{2}$Te$_{4}$, we approximate that $S_A\approx 1.47$ and $S_B\approx 0.32$. From fitting to the peak centers, we get the strength of the FM interaction $\mathcal{J}_0=-0.075(1)$~meV, and AFM interaction $\mathcal{J}'=0.75(3)$~meV. The calculated dispersions are shown in Fig.~\ref{fig:model}(c) and (d). At $\bm{q}=0$, spin wave mode $E_1$ goes to zero, which we can interpret as the acoustic spin wave formed mainly by the precession of Mn spins in the central Mn$_{\text{Mn}}$ layer. Optic mode $E_2$ gives a finite energy gap at $\bm{q}=0$, corresponding to the observed $\Delta' \approx 1.9$~meV, which results from the Mn-Sb anti-site mixing. Also, the flat-optic mode, $E_3$, arises from the top and bottom layers precessing out of phase with each other. Notice that $\mathcal{J}'$ is about 10 times strong than the dominant interaction $\mathcal{J}_0$ in MnBi$_2$Te$_4$~\cite{BLi2020,BLi2021} and MnBi$_4$Te$_7$~\cite{BLi2025}, in agreement with the value estimated from magnetization in high field~\cite{Lai2021}. 

\section*{References}
\label{sec:references}

\bibliography{reference.bib}

@article{kanec2005,
    title = {{Quantum Spin Hall Effect in Graphene}},
    author = {Kane, C. L. and Mele, E. J.},
    journal = {Phys. Rev. Lett.},
    volume = {95},
    issue = {22},
    pages = {226801},
    numpages = {4},
    year = {2005},
    month = {Nov},
    publisher = {American Physical Society},
    doi = {10.1103/PhysRevLett.95.226801},
    url = {https://link.aps.org/doi/10.1103/PhysRevLett.95.226801}
}

@article{kane2005,
    title = {{${Z}_{2}$ Topological Order and the Quantum Spin Hall Effect}},
    author = {Kane, C. L. and Mele, E. J.},
    journal = {Phys. Rev. Lett.},
    volume = {95},
    issue = {14},
    pages = {146802},
    numpages = {4},
    year = {2005},
    month = {Sep},
    publisher = {American Physical Society},
    doi = {10.1103/PhysRevLett.95.146802},
    url = {https://link.aps.org/doi/10.1103/PhysRevLett.95.146802}
}

@article{moore2007,
    title = {Topological invariants of time-reversal-invariant band structures},
    author = {Moore, J. E. and Balents, L.},
    journal = {Phys. Rev. B},
    volume = {75},
    issue = {12},
    pages = {121306},
    numpages = {4},
    year = {2007},
    month = {Mar},
    publisher = {American Physical Society},
    doi = {10.1103/PhysRevB.75.121306},
    url = {https://link.aps.org/doi/10.1103/PhysRevB.75.121306}
}

@article{zhang2009top,
    title={{Topological insulators in ${\mathrm{Bi}}_{2}{\mathrm{Se}}_{3}$, ${\mathrm{Bi}}_{2}{\mathrm{Te}}_{3}$ and ${\mathrm{Sb}}_{2}{\mathrm{Te}}_{3}$ with a single Dirac cone on the surface}},
    author={Zhang, Haijun and Liu, Chao-Xing and Qi, Xiao-Liang and Dai, Xi and Fang, Zhong and Zhang, Shou-Cheng},
    journal={Nat. Phys.},
    volume={5},
    number={6},
    pages={438--442},
    year={2009},
    publisher={Nature Publishing Group UK London},
    doi = {10.1038/nphys1270}
}

@article{yu2010quantized,
    title={{Quantized anomalous Hall effect in magnetic topological insulators}},
    author={Yu, Rui and Zhang, Wei and Zhang, Hai-Jun and Zhang, Shou-Cheng and Dai, Xi and Fang, Zhong},
    journal={Science},
    volume={329},
    number={5987},
    pages={61--64},
    year={2010},
    publisher={American Association for the Advancement of Science},
    doi = {10.1126/science.1187485}
}

@article{li2010dynamical,
    title={Dynamical axion field in topological magnetic insulators},
    author={Li, Rundong and Wang, Jing and Qi, Xiao-Liang and Zhang, Shou-Cheng},
    journal={Nat. Phys.},
    volume={6},
    number={4},
    pages={284--288},
    year={2010},
    publisher={Nature Publishing Group UK London},
    doi = {10.1038/nphys1534}
}

@article{chang2016quantum,
    title={{Quantum anomalous Hall effect in time-reversal-symmetry breaking topological insulators}},
    author={Chang, Cui-Zu and Li, Mingda},
    journal={J. Phys.: Condens. Matter},
    volume={28},
    number={12},
    pages={123002},
    year={2016},
    publisher={IOP Publishing},
    doi = {10.1088/0953-8984/28/12/123002}
}

@article{2017tailoring,
    title={Tailoring tricolor structure of magnetic topological insulator for robust axion insulator},
    author={Mogi, Masataka and Kawamura, Minoru and Tsukazaki, Atsushi and Yoshimi, Ryutaro and Takahashi, Kei S and Kawasaki, Masashi and Tokura, Yoshinori},
    journal={Sci. Adv.},
    volume={3},
    number={10},
    pages={eaao1669},
    year={2017},
    publisher={American Association for the Advancement of Science},
    doi = {10.1126/sciadv.aao1669}
}

@article{Mogi22,
    author = {Mogi, M. and Okamura, Y. and Kawamura, M. and Yoshimi, R. and Yasuda, K. and Tsukazaki, A. and Takahashi, K. S. and Morimoto, T. and Nagaosa, N. and Kawasaki, M. and Takahashi, Y. and Tokura, Y.},
    title = {Experimental signature of the parity anomaly in a semi-magnetic topological insulator},
    journal = {Nat. Phys.},
    volume = {18},
    number = {4},
    pages = {390-394},
    ISSN = {1745-2481},
    DOI = {10.1038/s41567-021-01490-y},
    url = {https://doi.org/10.1038/s41567-021-01490-y},
    year = {2022},
    type = {Journal Article}
}

@article{bernevig2022,
    title={Progress and prospects in magnetic topological materials},
    author={Bernevig, B Andrei and Felser, Claudia and Beidenkopf, Haim},
    journal={Nature},
    volume={603},
    number={7899},
    pages={41--51},
    year={2022},
    publisher={Nature Publishing Group UK London},
    doi = {10.1038/s41586-021-04105-x}
}

@article{Otrokov2019,
    title={Prediction and observation of an antiferromagnetic topological insulator},
    author={Otrokov, Mikhail M and Klimovskikh, Ilya I and Bentmann, Hendrik and Estyunin, D and Zeugner, Alexander and Aliev, Ziya S and Ga{\ss}, Sebastian and Wolter, AUB and Koroleva, AV and Shikin, Alexander M and Hoffmann, M. and Rusinov, I. P. and Vyazovskaya, A. Yu. and Eremeev, S. V. and Koroteev, Yu. M. and Kuznetsov, V. M. and Freyse, F. and Sánchez-Barriga, J. and Amiraslanov, I. R. and Babanly, M. B. and Mamedov, N. T. and Abdullayev, N. A. and Zverev, V. N. and Alfonsov, A. and Kataev, V. and Büchner, B. and Schwier, E. F. and Kumar, S. and Kimura, A. and Petaccia, L. and Di Santo, G. and Vidal, R. C. and Schatz, S. and Kißner, K. and Ünzelmann, M. and Min, C. H. and  Moser, Simon and Peixoto, T. R. F. and Reinert, F. and  Ernst, A. and Echenique, P. M. and Isaeva, A. and Chulkov, E. V.},
    journal={Nature},
    volume={576},
    number={7787},
    pages={416--422},
    year={2019},
    publisher={Nature Publishing Group UK London},
    doi = {10.1038/s41586-019-1840-9},
    url = {https://doi.org/10.1038/s41586-019-1840-9}
}

@article{Zhang19,
    author = {Zhang, Dongqin and Shi, Minji and Zhu, Tongshuai and Xing, Dingyu and Zhang, Haijun and Wang, Jing},
    title = {{Topological Axion States in the Magnetic Insulator ${\mathrm{MnBi}}_{2}{\mathrm{Te}}_{4}$ with the Quantized Magnetoelectric Effect}},
    journal = {Phys. Rev. Lett.},
    volume = {122},
    number = {20},
    pages = {206401},
    DOI = {10.1103/PhysRevLett.122.206401},
    url = {https://journals.aps.org/prl/pdf/10.1103/PhysRevLett.122.206401},
    year = {2019},
    type = {Journal Article}
}

@article{JLi2019,
    author = {Jiaheng Li and Yang Li and Shiqiao Du and Zun Wang and Bing-Lin Gu and Shou-Cheng Zhang and Ke He and Wenhui Duan  and Yong Xu },
    title = {{Intrinsic magnetic topological insulators in van der Waals layered ${\mathrm{MnBi}}_{2}{\mathrm{Te}}_{4}$-family materials}},
    journal = {Sci. Adv.},
    volume = {5},
    number = {6},
    pages = {eaaw5685},
    year = {2019},
    doi = {10.1126/sciadv.aaw5685},
    URL = {https://www.science.org/doi/abs/10.1126/sciadv.aaw5685}
}

@article{Gong19,
    author = {Gong, Yan and Guo, Jingwen and Li, Jiaheng and Zhu, Kejing and Liao, Menghan and Liu, Xiaozhi and Zhang, Qinghua and Gu, Lin and Tang, Lin and Feng, Xiao and Zhang, Ding and Li, Wei and Song, Canli and Wang, Lili and Yu, Pu and Chen, Xi and Wang, Yayu and Yao, Hong and Duan, Wenhui and Xu, Yong and Zhang, Shou-Cheng and Ma, Xucun and Xue, Qi-Kun and He, Ke},
    title = {{Experimental Realization of an Intrinsic Magnetic Topological Insulator}},
    journal = {Chin. Phys. Lett.},
    volume = {36},
    number = {7},
    pages = {076801},
    ISSN = {0256-307X1741-3540},
    DOI = {10.1088/0256-307x/36/7/076801},
    url = {https://iopscience.iop.org/article/10.1088/0256-307X/36/7/076801/pdf},
    year = {2019},
    type = {Journal Article}
}

@article{Li2024review,
    author = {Li, Shuai and Liu, Tianyu and Liu, Chang and Wang, Yayu and Lu, Hai-Zhou and Xie, X C},
    title = {{Progress on the antiferromagnetic topological insulator ${\mathrm{MnBi}}_{2}{\mathrm{Te}}_{4}$}},
    journal = {Nat. Sci. Rev.},
    volume = {11},
    number = {2},
    pages = {nwac296},
    ISSN = {2095-5138},
    doi = {10.1093/nsr/nwac296},
    year = {2024},
    type = {Journal Article}
}

@article{Vyazovskaya25,
    author = {Vyazovskaya, Alexandra Yu and Bosnar, Mihovil and Chulkov, Evgueni V. and Otrokov, Mikhail M.},
    title = {{Intrinsic magnetic topological insulators of the ${\mathrm{MnBi}}_{2}{\mathrm{Te}}_{4}$ family}},
    journal = {Commun. Mater.},
    volume = {6},
    number = {1},
    pages = {88},
    ISSN = {2662-4443},
    DOI = {10.1038/s43246-025-00794-3},
    url = {https://doi.org/10.1038/s43246-025-00794-3},
    year = {2025},
    type = {Journal Article}
}

@article{Deng2020,
    title = {{Quantum anomalous Hall effect in intrinsic magnetic topological insulator ${\mathrm{MnBi}}_{2}{\mathrm{Te}}_{4}$}},
    author = {Yujun Deng and Yijun Yu and Meng Zhu Shi  and Zhongxun Guo and Zihan Xu and Jing Wang and Xian Hui Chen and Yuanbo Zhang},
    journal = {Science},
    volume = {367},
    number = {6480},
    pages = {895-900},
    year = {2020},
    doi = {10.1126/science.aax8156},
    URL = {https://www.science.org/doi/abs/10.1126/science.aax8156}
}

@article{Liu20,
    author = {Liu, Chang and Wang, Yongchao and Li, Hao and Wu, Yang and Li, Yaoxin and Li, Jiaheng and He, Ke and Xu, Yong and Zhang, Jinsong and Wang, Yayu},
    title = {{Robust axion insulator and Chern insulator phases in a two-dimensional antiferromagnetic topological insulator}},
    journal = {Nat. Mater.},
    volume = {19},
    number = {5},
    pages = {522-527},
    ISSN = {1476-4660},
    DOI = {10.1038/s41563-019-0573-3},
    url = {https://doi.org/10.1038/s41563-019-0573-3},
    year = {2020},
    type = {Journal Article}
}

@article{Lin22,
    author = {Lin, Weiyan and Feng, Yang and Wang, Yongchao and Zhu, Jinjiang and Lian, Zichen and Zhang, Huanyu and Li, Hao and Wu, Yang and Liu, Chang and Wang, Yihua and Zhang, Jinsong and Wang, Yayu and Chen, Chui-Zhen and Zhou, Xiaodong and Shen, Jian},
    title = {{Direct visualization of edge state in even-layer ${\mathrm{MnBi}}_{2}{\mathrm{Te}}_{4}$ at zero magnetic field}},
    journal = {Nat. Commun.},
    volume = {13},
    number = {1},
    pages = {7714},
    ISSN = {2041-1723},
    DOI = {10.1038/s41467-022-35482-0},
    url = {https://doi.org/10.1038/s41467-022-35482-0},
    year = {2022},
    type = {Journal Article}
}

@article{qiu2025observation,
    title={{Observation of the axion quasiparticle in 2D ${\mathrm{MnBi}}_{2}{\mathrm{Te}}_{4}$}},
    author={Qiu, Jian-Xiang and Ghosh, Barun and Sch{\"u}tte-Engel, Jan and Qian, Tiema and Smith, Michael and Yao, Yueh-Ting and Ahn, Junyeong and Liu, Yu-Fei and Gao, Anyuan and Tzschaschel, Christian and Li, Houchen and Petrides, Ioannis and Bérubé, Damien and Dinh, Thao and Huang, Tianye and Liebman, Olivia and Been, Emily M. and Blawat, Joanna M. and Watanabe, Kenji and Taniguchi, Takashi and Fong, Kin Chung and Lin, Hsin and Orth, Peter P. and Narang, Prineha and Felser, Claudia and Chang, Tay-Rong and McDonald, Ross and McQueeney, Robert J. and Bansil, Arun and Martin, Ivar and Ni, Ni and Ma, Qiong and Marsh, David J. E. and Vishwanath, Ashvin and Xu, Su-Yang},
    journal={Nature},
    volume={641},
    number={8061},
    pages={62--69},
    year={2025},
    publisher={Nature Publishing Group UK London},
    doi = {10.1038/s41586-025-08862-x}
}

@article{Gao21,
    author = {Gao, Anyuan and Liu, Yu-Fei and Hu, Chaowei and Qiu, Jian-Xiang and Tzschaschel, Christian and Ghosh, Barun and Ho, Sheng-Chin and Bérubé, Damien and Chen, Rui and Sun, Haipeng and Zhang, Zhaowei and Zhang, Xin-Yue and Wang, Yu-Xuan and Wang, Naizhou and Huang, Zumeng and Felser, Claudia and Agarwal, Amit and Ding, Thomas and Tien, Hung-Ju and Akey, Austin and Gardener, Jules and Singh, Bahadur and Watanabe, Kenji and Taniguchi, Takashi and Burch, Kenneth S. and Bell, David C. and Zhou, Brian B. and Gao, Weibo and Lu, Hai-Zhou and Bansil, Arun and Lin, Hsin and Chang, Tay-Rong and Fu, Liang and Ma, Qiong and Ni, Ni and Xu, Su-Yang},
    title = {Layer Hall effect in a 2D topological axion antiferromagnet},
    journal = {Nature},
    volume = {595},
    number = {7868},
    pages = {521-525},
    ISSN = {1476-4687},
    DOI = {10.1038/s41586-021-03679-w},
    url = {https://doi.org/10.1038/s41586-021-03679-w},
    year = {2021},
    type = {Journal Article}
}

@article{tokura2019magnetic,
    title={{Magnetic topological insulators}},
    author={Tokura, Yoshinori and Yasuda, Kenji and Tsukazaki, Atsushi},
    journal={Nat. Rev. Phys.},
    volume={1},
    number={2},
    pages={126--143},
    year={2019},
    publisher={Nature Publishing Group UK London},
    doi = {10.1038/s42254-018-0011-5}
}

@article{YSHor2010,
    title = {{Development of ferromagnetism in the doped topological insulator ${\text{Bi}}_{2\ensuremath{-}x}{\text{Mn}}_{x}{\text{Te}}_{3}$}},
    author = {Hor, Y. S. and Roushan, P. and Beidenkopf, H. and Seo, J. and Qu, D. and Checkelsky, J. G. and Wray, L. A. and Hsieh, D. and Xia, Y. and Xu, S.-Y. and Qian, D. and Hasan, M. Z. and Ong, N. P. and Yazdani, A. and Cava, R. J.},
    journal = {Phys. Rev. B},
    volume = {81},
    issue = {19},
    pages = {195203},
    numpages = {7},
    year = {2010},
    month = {May},
    publisher = {American Physical Society},
    doi = {10.1103/PhysRevB.81.195203},
    url = {https://link.aps.org/doi/10.1103/PhysRevB.81.195203}
}

@article{ZhangPRB2013,
    title = {{Stability, electronic, and magnetic properties of the magnetically doped topological insulators ${\text{Bi}}_{2}$Se${}_{3}$, Bi${}_{2}$Te${}_{3}$, and Sb${}_{2}$Te${}_{3}$}},
    author = {Zhang, Jian-Min and Ming, Wenmei and Huang, Zhigao and Liu, Gui-Bin and Kou, Xufeng and Fan, Yabin and Wang, Kang L. and Yao, Yugui},
    journal = {Phys. Rev. B},
    volume = {88},
    issue = {23},
    pages = {235131},
    numpages = {9},
    year = {2013},
    month = {Dec},
    publisher = {American Physical Society},
    doi = {10.1103/PhysRevB.88.235131},
    url = {https://link.aps.org/doi/10.1103/PhysRevB.88.235131}
}

@article{islam2023role,
    title={Role of magnetic defects in tuning ground states of magnetic topological insulators},
    author={Islam, Farhan and Lee, Yongbin and Pajerowski, Daniel M and Oh, JinSu and Tian, Wei and Zhou, Lin and Yan, Jiaqiang and Ke, Liqin and McQueeney, Robert J and Vaknin, David},
    journal={Adv. Mater.},
    volume={35},
    number={21},
    pages={2209951},
    year={2023},
    publisher={Wiley Online Library},
    doi = {10.1002/adma.202209951}
}

@article{Lai2021,
    title = {{Defect-driven ferrimagnetism and hidden magnetization in ${\mathrm{MnBi}}_{2}{\mathrm{Te}}_{4}$}},
    author = {Lai, You and Ke, Liqin and Yan, Jiaqiang and McDonald, Ross D. and McQueeney, Robert J.},
    journal = {Phys. Rev. B},
    volume = {103},
    issue = {18},
    pages = {184429},
    numpages = {8},
    year = {2021},
    month = {May},
    publisher = {American Physical Society},
    doi = {10.1103/PhysRevB.103.184429},
    url = {https://link.aps.org/doi/10.1103/PhysRevB.103.184429}
}

@article{Liu2021,
    title = {{Site Mixing for Engineering Magnetic Topological Insulators}},
    author = {Liu, Yaohua and Wang, Lin-Lin and Zheng, Qiang and Huang, Zengle and Wang, Xiaoping and Chi, Miaofang and Wu, Yan and Chakoumakos, Bryan C. and McGuire, Michael A. and Sales, Brian C. and Wu, Weida and Yan, Jiaqiang},
    journal = {Phys. Rev. X},
    volume = {11},
    issue = {2},
    pages = {021033},
    numpages = {12},
    year = {2021},
    month = {May},
    publisher = {American Physical Society},
    doi = {10.1103/PhysRevX.11.021033},
    url = {https://link.aps.org/doi/10.1103/PhysRevX.11.021033}
}

@article{garnica2022,
    title={{Native point defects and their implications for the Dirac point gap at ${\mathrm{MnBi}}_{2}{\mathrm{Te}}_{4}$ (0001)}},
    author={Garnica, Manuela and Otrokov, Mikhail M and Aguilar, P Casado and Klimovskikh, Ilya I and Estyunin, Dmitry and Aliev, Ziya S and Amiraslanov, Imamaddin R and Abdullayev, Nadir A and Zverev, Vladimir N and Babanly, Mahammad B and Mamedov, N. T. and Shikin, A. M. and Arnau, A. and de Parga, A. L. Vázquez and Chulkov, E. V. and Miranda, R.},
    journal={npj Quant. Mater.},
    volume={7},
    number={1},
    pages={7},
    year={2022},
    publisher={Nature Publishing Group UK London},
    URL = {https://doi.org/10.1038/s41535-021-00414-6}
}

@article{TaitoPRB2019,
    title = {{Realization of interlayer ferromagnetic interaction in $\mathrm{MnS}{\mathrm{b}}_{2}\mathrm{T}{\mathrm{e}}_{4}$ toward the magnetic Weyl semimetal state}},
    author = {Murakami, Taito and Nambu, Yusuke and Koretsune, Takashi and Xiangyu, Gu and Yamamoto, Takafumi and Brown, Craig M. and Kageyama, Hiroshi},
    journal = {Phys. Rev. B},
    volume = {100},
    issue = {19},
    pages = {195103},
    numpages = {6},
    year = {2019},
    month = {Nov},
    publisher = {American Physical Society},
    doi = {10.1103/PhysRevB.100.195103},
    url = {https://link.aps.org/doi/10.1103/PhysRevB.100.195103}
}

@article{Hu24,
    author = {Hu, Xinmeng and He, Xinyi and Guo, Zhilin and Kamiya, Toshio and Wu, Jiazhen},
    title = {{Antisite-Defects Control of Magnetic Properties in ${\mathrm{MnSb}}_{2}{\mathrm{Te}}_{4}$}},
    journal = {ACS Nano},
    volume = {18},
    number = {1},
    pages = {738-749},
    ISSN = {1936-0851},
    DOI = {10.1021/acsnano.3c09064},
    url = {https://doi.org/10.1021/acsnano.3c09064},
    year = {2024},
    type = {Journal Article}
}

@article{Sahoo24,
    author = {Sahoo, Manaswini and Onuorah, Ifeanyi John and Folkers, Laura Christina and Kochetkova, Ekaterina and Chulkov, Evgueni V. and Otrokov, Mikhail M. and Aliev, Ziya S. and Amiraslanov, Imamaddin R. and Wolter, Anja U. B. and Büchner, Bernd and Corredor, Laura Teresa and Wang, Chennan and Salman, Zaher and Isaeva, Anna and De Renzi, Roberto and Allodi, Giuseppe},
    title = {{Ubiquitous Order-Disorder Transition in the Mn Antisite Sublattice of the  (${\mathrm{MnBi}}_{2}{\mathrm{Te}}_{4}$)(${\mathrm{Bi}}_{2}{\mathrm{Te}}_{3})_{\mathrm{n}}$ Magnetic Topological Insulators}},
    journal = {Adv. Sci.},
    volume = {11},
    number = {34},
    pages = {2402753},
    ISSN = {2198-3844},
    DOI = {10.1002/advs.202402753},
    url = {https://advanced.onlinelibrary.wiley.com/doi/abs/10.1002/advs.202402753},
    year = {2024},
    type = {Journal Article}
}

@article{Riberolles21,
    author = {Riberolles, S. X. M. and Zhang, Q. and Gordon, Elijah and Butch, N. P. and Ke, Liqin and Yan, J. Q. and McQueeney, R. J.},
    title = {{Evolution of magnetic interactions in Sb-substituted ${\mathrm{MnBi}}_{2}{\mathrm{Te}}_{4}$}},
    journal = {Phys. Rev. B},
    volume = {104},
    number = {6},
    pages = {064401},
    DOI = {10.1103/PhysRevB.104.064401},
    url = {https://link.aps.org/doi/10.1103/PhysRevB.104.064401},
    year = {2021},
    type = {Journal Article}
}

@article{Yan19,
    author = {Yan, J. Q. and Okamoto, S. and McGuire, M. A. and May, A. F. and McQueeney, R. J. and Sales, B. C.},
    title = {{Evolution of structural, magnetic, and transport properties in  ${\mathrm{MnBi}}_{2-x}{\mathrm{Sb}}_{x}{\mathrm{Te}}_{4}$}},
    journal = {Phys. Rev. B},
    volume = {100},
    number = {10},
    pages = {104409},
    DOI = {10.1103/PhysRevB.100.104409},
    url = {https://link.aps.org/doi/10.1103/PhysRevB.100.104409},
    year = {2019},
    type = {Journal Article}
}

@article{Hu21,
    author = {Hu, Chaowei and Lien, Shang-Wei and Feng, Erxi and Mackey, Scott and Tien, Hung-Ju and Mazin, Igor I. and Cao, Huibo and Chang, Tay-Rong and Ni, Ni},
    title = {{Tuning magnetism and band topology through antisite defects in Sb-doped ${\mathrm{MnBi}}_{4}{\mathrm{Te}}_{7}$}},
    journal = {Phys. Rev. B},
    volume = {104},
    number = {5},
    pages = {054422},
    DOI = {10.1103/PhysRevB.104.054422},
    url = {https://link.aps.org/doi/10.1103/PhysRevB.104.054422},
    year = {2021},
    type = {Journal Article}
}

@article{Pham2019,
    title = {Quantum material topology via defect engineering},
    author = {Pham, Anh and Ganesh, P.},
    journal = {Phys. Rev. B},
    volume = {100},
    issue = {24},
    pages = {241110(R)},
    numpages = {6},
    year = {2019},
    month = {Dec},
    publisher = {American Physical Society},
    doi = {10.1103/PhysRevB.100.241110},
    url = {https://link.aps.org/doi/10.1103/PhysRevB.100.241110}
}

@article{Du2021,
    author = {Du, Mao-Hua and Yan, Jiaqiang and Cooper, Valentino R. and Eisenbach, Markus},
    title = {{Tuning Fermi Levels in Intrinsic Antiferromagnetic Topological Insulators ${\mathrm{MnBi}}_{2}{\mathrm{Te}}_{4}$ and ${\mathrm{MnBi}}_{4}{\mathrm{Te}}_{7}$ by Defect Engineering and Chemical Doping}},
    journal = {Adv. Funct. Mater.},
    volume = {31},
    number = {3},
    pages = {2006516},
    year = {2021},
    doi = {https://doi.org/10.1002/adfm.202006516},
    url = {https://advanced.onlinelibrary.wiley.com/doi/abs/10.1002/adfm.202006516}
}

@article{chen2025defect,
    title = {{Defect Engineering for Stabilizing Magnetic and Topological Properties in Mn $({\mathrm{Bi}}_{1-x}{\mathrm{Sb}}_{x})_{2}{\mathrm{Te}}_{4}$}},
    author = {Chen, Haonan and Wang, Jiayu and Li, Huayao and Duan, Xunkai and Wang, Yuxiang and Xu, Zixuan and Xia, Yingchao and He, Wenhao and Jia, Zehao and Cao, Xiangyu and Mou, Yicheng and Jiang, Xiangyu and Gu, Jiaming and Leng, Pengliang and Zhu, Fengfeng and Zheng, Changlin and Yuan, Xiang and Xiu, Faxian and Zhou, Tong and Miao, Lin and Zhang, Cheng},
    journal = {Nat. Commun.},
    volume = {17},
    pages = {1029},
    year = {2025},
    publisher={Nature Publishing Group UK London},
    doi = {10.1038/s41467-025-67774-6},
    url = {https://doi.org/10.1038/s41467-025-67774-6}
}

@article{Enk025atomic,
    title={{Atomic engineering of intrinsic permanent magnetism in MnBi}},
    author={Enkhtur, Uranbaigal and Odkhuu, Dorj},
    journal={Sci. Rep.},
    volume={15},
    number={1},
    pages={36792},
    year={2025},
    publisher={Nature Publishing Group UK London},
    doi = {10.1038/s41598-025-20764-6},
    url = {https://doi.org/10.1038/s41598-025-20764-6}
}

@article{BLi2021,
    title = {{Quasi-two-dimensional ferromagnetism and anisotropic interlayer couplings in the magnetic topological insulator ${\mathrm{MnBi}}_{2}{\mathrm{Te}}_{4}$}},
    author = {Li, Bing and Pajerowski, D. M. and Riberolles, S. X. M. and Ke, Liqin and Yan, J.-Q. and McQueeney, R. J.},
    journal = {Phys. Rev. B},
    volume = {104},
    issue = {22},
    pages = {L220402},
    numpages = {6},
    year = {2021},
    month = {Dec},
    publisher = {American Physical Society},
    doi = {10.1103/PhysRevB.104.L220402},
    url = {https://link.aps.org/doi/10.1103/PhysRevB.104.L220402}
}

@article{Yan2019,
    title = {{Crystal growth and magnetic structure of ${\mathrm{MnBi}}_{2}{\mathrm{Te}}_{4}$}},
    author = {Yan, J.-Q. and Zhang, Q. and Heitmann, T. and Huang, Zengle and Chen, K. Y. and Cheng, J.-G. and Wu, Weida and Vaknin, D. and Sales, B. C. and McQueeney, R. J.},
    journal = {Phys. Rev. Mater.},
    volume = {3},
    issue = {6},
    pages = {064202},
    numpages = {8},
    year = {2019},
    month = {Jun},
    publisher = {American Physical Society},
    doi = {10.1103/PhysRevMaterials.3.064202},
    url = {https://link.aps.org/doi/10.1103/PhysRevMaterials.3.064202}
}

@article{BLi2020,
    title = {{Competing Magnetic Interactions in the Antiferromagnetic Topological Insulator ${\mathrm{MnBi}}_{2}{\mathrm{Te}}_{4}$}},
    author = {Li, Bing and Yan, J.-Q. and Pajerowski, D. M. and Gordon, Elijah and Nedi\ifmmode \acute{c}\else \'{c}\fi{}, A.-M. and Sizyuk, Y. and Ke, Liqin and Orth, P. P. and Vaknin, D. and McQueeney, R. J.},
    journal = {Phys. Rev. Lett.},
    volume = {124},
    issue = {16},
    pages = {167204},
    numpages = {6},
    year = {2020},
    month = {Apr},
    publisher = {American Physical Society},
    doi = {10.1103/PhysRevLett.124.167204},
    url = {https://link.aps.org/doi/10.1103/PhysRevLett.124.167204}
}

@article{Furrer2013,
    title = {Magnetic cluster excitations},
    author = {Furrer, Albert and Waldmann, Oliver},
    journal = {Rev. Mod. Phys.},
    volume = {85},
    issue = {1},
    pages = {367--420},
    numpages = {0},
    year = {2013},
    month = {Mar},
    publisher = {American Physical Society},
    doi = {10.1103/RevModPhys.85.367},
    url = {https://link.aps.org/doi/10.1103/RevModPhys.85.367}
}

@article{Vaknin20,
    author = {Vaknin, D. and Pakhira, Santanu and Schlagel, D. and Islam, F. and Zhang, Jianhua and Pajerowski, D. M. and Wang, C. Z. and Johnston, D. C. and McQueeney, R. J.},
    title = {{Localized singlets and ferromagnetic fluctuations in the dilute magnetic topological insulator ${\mathrm{Sn}}_{0.95}{\mathrm{Mn}}_{0.05}{\mathrm{Te}}$}},
    journal = {Phys. Rev. B},
    volume = {101},
    number = {14},
    pages = {140406},
    DOI = {10.1103/PhysRevB.101.140406},
    url = {https://link.aps.org/doi/10.1103/PhysRevB.101.140406},
    year = {2020},
    type = {Journal Article}
}

@article{BLi2025,
    title = {{Role of nonmagnetic spacers in the magnetic interactions of antiferromagnetic topological insulators ${\mathrm{MnBi}}_{4}{\mathrm{Te}}_{7}$ and ${\mathrm{MnBi}}_{2}{\mathrm{Te}}_{4}$}},
    author = {Li, Bing and Pajerowski, D. M. and Yan, J.-Q. and McQueeney, R. J.},
    journal = {Phys. Rev. B},
    volume = {111},
    issue = {6},
    pages = {064418},
    numpages = {6},
    year = {2025},
    month = {Feb},
    publisher = {American Physical Society},
    doi = {10.1103/PhysRevB.111.064418},
    url = {https://link.aps.org/doi/10.1103/PhysRevB.111.064418}
}

@article{Goodenough,
      title = {{Theory of the Role of Covalence in the Perovskite-Type Manganites $[\mathrm{La}, M(\mathrm{II})]\mathrm{Mn}{\mathrm{O}}_{3}$}},
      author = {Goodenough, John B.},
      journal = {Phys. Rev.},
      volume = {100},
      issue = {2},
      pages = {564--573},
      numpages = {0},
      year = {1955},
      month = {Oct},
      publisher = {American Physical Society},
      doi = {10.1103/PhysRev.100.564},
      url = {https://link.aps.org/doi/10.1103/PhysRev.100.564}
}

@article{Kanamori,
    title = {Superexchange interaction and symmetry properties of electron orbitals},
    journal = {J. Phys.  Chem. Solids},
    volume = {10},
    number = {2},
    pages = {87-98},
    year = {1959},
    issn = {0022-3697},
    doi = {https://doi.org/10.1016/0022-3697(59)90061-7},
    url = {https://www.sciencedirect.com/science/article/pii/0022369759900617},
    author = {Junjiro Kanamori},
}

@article{Tyler_PRM,
    title = {{Vacancy-tuned magnetism in ${\mathrm{LaMn}}_{x}{\mathrm{Sb}}_{2}$}},
    author = {Slade, Tyler J. and Sapkota, Aashish and Wilde, John M. and Zhang, Qiang and Wang, Lin-Lin and Lapidus, Saul H. and Schmidt, Juan and Heitmann, Thomas and Bud'ko, Sergey L. and Canfield, Paul C.},
    journal = {Phys. Rev. Mater.},
    volume = {7},
    issue = {11},
    pages = {114203},
    numpages = {25},
    year = {2023},
    month = {Nov},
    publisher = {American Physical Society},
    doi = {10.1103/PhysRevMaterials.7.114203},
    url = {https://link.aps.org/doi/10.1103/PhysRevMaterials.7.114203}
}

@article{Dahlbom2025,
    title = {{Sunny.jl: A Julia Package for Spin Dynamics}},
    author = {Dahlbom, David and Zhang, Hao and Miles, Cole and Quinn, Sam and Niraula, Alin and Thipe, Bhushan and Wilson, Matthew and Matin, Sakib and Mankad, Het and Hahn, Steven and Pajerowski, Daniel and Johnston, Steve and Wang, Zhentao and Lane, Harry and Li, Ying Wai and Bai, Xiaojian and Mourigal, Martin and Batista, Cristian D. and Barros, Kipton},
    journal = {J. Open Source Softw.}, 
    volume = {10},
    number = {116},
    pages = {8138},
    year = {2025},
    publisher = {The Open Journal},
    doi = {10.21105/joss.08138},
    url = {https://doi.org/10.21105/joss.08138}
}

@misc{IPTS,
    author = {Jaishi, Dhurba and Li, Bing and McQueeney, R. J. and Pajerowski, D. M.},
    title = {{Spin dynamics in dilute ${\mathrm{MnSb}}_{2}{\mathrm{Te}}_{4}$}},
    howpublished  = {{[Link Here]}},
}
\end{document}